\documentclass[aps,pra,twocolumn,showpacs,showkeys,reprint]{revtex4-1} 
\usepackage{graphics} 
\usepackage{graphicx} 
\usepackage{amsmath} 

\begin{document} 

\title{Spin liquid state in an inhomogeneous periodic Anderson model} 
\author{R. C. Caro} 
\author{R. Franco} 
\author{J. Silva-Valencia} 
\email{jsilvav@unal.edu.co} 
\affiliation{Departamento de F\'{\i}sica, Universidad Nacional de Colombia, A. A. 5997 Bogot\'a, Colombia.} 

\date{\today} 

\begin{abstract} 
We studied the ground state of alkaline-earth-metal atoms confined in one-dimensional optical lattices with an effective hybridization generated by a 
suitable laser field. This system is modeled by the periodic Anderson model plus a quadratic confining potential, and we adopted the density-matrix 
renormalization group to calculate its ground state. We found a one-to-one correspondence between the local variance, the local von Neumann entropy, 
and the on-site spin-spin correlation. For low global densities, we observed the formation of local singlets between delocalized and localized atoms 
and found Kondo spin liquid domains that can be tuned with the confining potential, the hybridization, and the local repulsion. Band insulator, 
metallic, phase separation, and Kondo spin liquid regions coexist in the ground state.

\end{abstract} 


\maketitle 

\section{\label{sec1}Introduction} 
Materials such as $CeAl_3$, $FeSi$, $YbCu_{2}Si_{2}$, and $UBe_{13}$, among others, are commonly known as heavy fermions, due to the enormous 
effective mass of their charge carriers. Since 1975, these materials have been intensely studied, exhibiting a variety of very interesting exotic 
states at very low temperatures,  such as superconductivity, spin-liquid or even metal-insulator 
coexistence~\cite{Grewe-CBook91,Wachter-CBook94,Coleman-CBook07,Gegenwar-NP08}. Up to today, it is believed that the physical properties of 
heavy fermions can be described by the competition between two fundamental interactions: the Kondo effect, or the partial shielding of the magnetic 
moments by conduction electrons , and the Ruderman-Kittel-Kosiya-Yosida (RKKY) interaction, or the long-range magnetic 
interaction between localized moments~\cite{Hewson-Book}. A magnetic ordering in the heavy fermion material is expected when the RKKY 
interaction has a higher value than the Kondo effect and a non-magnetic state takes place in the opposite case. However, quite interesting 
physical behavior happens when a balance of the competition between these two interactions occurs. To explain the experimental results, two-orbital 
models have been considered, the periodic Anderson model and the Kondo lattice model, the former being the most general, taking into account the 
hybridization between orbitals~\cite{Tsunetsugu-RMP97,Shibata-JPCM99}. The above models have allowed us to understand some experimental results for  
$FeSi$, among others; however, the experimental spin gap and charge gap ratio for $CeRhAs$ cannot be explained with these models~\cite{Hagymasi-PRB14}. 
The possibility of exploring the relevance of the diverse processes described by the models and controlling the strength of the parameters is remote 
for the materials, due to impurities, defects, and other difficulties.\par 
Ultracold atom setups have become a laboratory for the condensed matter field. With them, several ideas and concepts have been tested and extended, 
in an environment that is clean and free of impurities and defects, with complete control over the parameters: tunneling, density, and interactions 
between atoms~\cite{Lewenstein-AP07,IBloch-08,IBloch-12}. Basic models as such as Bose- and Fermi-Hubbard models have been emulated confining atoms 
in optical lattices, 
recovering known results and finding new phenomena by exploring the models~\cite{Greiner-N02a,Jordens-N08}. At the beginning of this decade, 
the possibility of studying the interplay between spin and charge degrees of freedom with alkaline-earth atoms confined in optical lattices was 
raised by Gorshkov \textit{et al.}, opening the door to studying $SU(N)$ Hubbard models and two-orbital models~\cite{Gorshkov-NP10}. Others theoretical 
proposals for studying the Kondo effect in optical lattices followed~\cite{Nishida-PRL13,Bauer-PRL13}, and several experimental efforts have allowed to 
estimating the value of two-orbital spin-exchange interactions between atoms in different optical lattices, which suggests the possibility of emulating 
the  Kondo lattice model in these fully tunable setups~\cite{Scazza-NP14,Zhang-S14,Cappellini-PRL15,Riegger-arxiv17}.\par
In 2005, Paredes \textit{et al.} proposed using an optical superlattice to realize a particular Anderson lattice model in which 
localized atoms interact with each other through a discretized set of delocalized levels, finding a magnetic phase with a long-range magnetic order 
between the localized atoms and a Kondo-singlet phase~\cite{Paredes-PRA05}.\par
Nakagawa and Kawakami suggested the use of a laser to couple the two internal states of alkaline-earth atoms and described the system in terms of a 
``hybridization quench'' problem of the Anderson lattice model. They showed the possibility to obtain a 
Kondo-singlet state with atoms confined in optical lattices, overcoming the drawback of the heat generated by the laser~\cite{Nakagawa-PRL15}.\par 
Recently, another theoretical proposal for emulating the periodic Anderson model in optical lattices was made by Y. Zhong \textit{et al.}~\cite{Zhong-FP17},
who suggested the use of a suitable laser field to generate transitions between the optical lattices that emulate the orbitals, achieving an effective 
hybridization, which is a key ingredient in the periodic Anderson model. They studied the periodic Anderson model under a harmonic potential in one 
and two dimensions using quantum Monte Carlo and dynamic mean-field theory, finding for the half-filling case metallic and insulating regions coexisting 
and a localized optical lattice with one atom per site around the middle of the trap. Also, their results suggest the formation of some singlets in the 
middle of the trap. One of the well-known facts about the two-orbital models is the spin liquid state, characterized by the formation of local 
singlets between carriers in different 
orbitals~\cite{Blankenbecler-PRL87,Ueda-PRL92,Nishino-PRB93,Tian-PRB94,Varma-PRB94,Guerrero-PRB95,Luo-PRB02,Bertussi-PRB11}. 
Taking into account the above, we decided to explore the formation of spin liquid regions inside of the  
trap as a function of the global density, the confining potential, the hybridization, and the local repulsion between the localized atoms. Using the 
density matrix renormalization group method to study the ground state of the periodic Anderson model with a harmonic potential, we observed the 
coexistence of metallic and insulating regions along the lattice for different densities. We identified a spin liquid region around the middle of the 
trap. Due to the confinement at the ends of the lattice, the system always exhibits band insulator regions (empty sites), but as the confinement 
increased, we observed around the middle of the trap from metallic to band insulator (with full sites) regions passing through the spin liquid region.
We also found that it is possible to tune the spin liquid region with the hybridization and the local repulsion between the localized atoms.\par 
The rest of the paper is organized as follows: in Section \ref{sec2} we discuss the periodic Anderson model with a harmonic potential and the known 
results in the limit cases. The ground state as a function of the confinement potential, the hybridization, and the local repulsion, taking into 
account different densities, are shown in Section \ref{sec3}. Finally, the conclusions are presented in Section \ref{sec4}.\par 
\begin{figure}[t]
\centering
\includegraphics[width=0.5\textwidth]{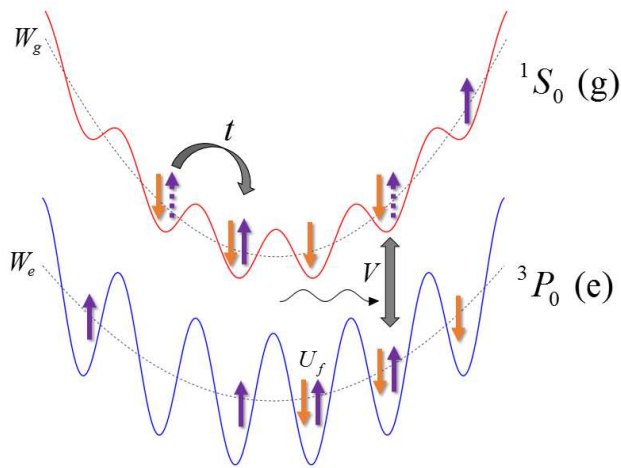}
\caption{Schematic of the setup related to Hamiltonian (\ref{HPAM}). $^{171}Yb$ atoms can be confined independently in two different optical 
lattices. The $g$ ($e$) atoms are itinerant (localized) and a hybridization between the optical lattices can be achieved by a suitable laser field. 
Due to the optical lattices, the atoms undergo a harmonic trapping potential, whose strength is the same for both types of atoms ($W_e=W_g=W$). $V$ 
represents the effective hybridization between the atoms, $U_f$ is the repulsion between the localized atoms, and $t$ is the hopping parameter of 
itinerant atoms. The nuclear spin degree of freedom of the atoms is represented by arrows.}
\label{fig1}
\end{figure} 
\section{\label{sec2} Model}
The experimental achievement of obtaining a degenerate gas of $^{171}Yb$ atoms has increased the chances of studying the charge and spin interplay 
in optical lattice setups~\cite{Taie-PRL10}, because these atoms have a nuclear spin of $I=1/2$ and emulate the electrons in materials. 
The existence of a long-lived metastable excited state $^{3}P_{0}$ ($\arrowvert e\rangle$)  coupled to the ground state $^{1}S_{0}$ 
($\arrowvert g\rangle$) via an ultranarrow doubly-forbidden transition allows studying the physics of two-orbital materials with alkaline-earth-like 
atoms, because it is possible to confine atoms independently in two different optical lattice potentials with the same 
periodicity~\cite{Daley-PRL08}. At low temperatures, the spin-changing collisions are prohibited and four different scattering lengths for the states 
$\arrowvert ee\rangle$,$\arrowvert gg\rangle$, and $\frac{1}{\sqrt{2}}(\arrowvert ge\rangle \pm \arrowvert eg\rangle)$ arise. In the experimental 
proposal by Zhong \textit{et al.}~\cite{Zhong-FP17}, a transition between atoms in different optical lattices 
($\arrowvert e\rangle \Leftrightarrow \arrowvert g\rangle$) by means of a suitable laser field is suggested, which generates an effective $g-e$ 
hybridization between the atoms in different optical lattices, in a way similar to the conduction and the local electron's hybridization in the heavy 
fermion materials. Tuning the interactions such that the $g$ atoms do not interact between them and locating the $e$ atoms in a deep optical 
lattice, we arrived at a system that emulates the periodic Anderson model (PAM) with atoms confined in optical lattices, whose Hamiltonian is given by:
\begin{eqnarray}\label{HPAM} 
\mathcal{H}=&-&t\sum_{i,\sigma}\left(c^{\dagger}_{i\sigma}c_{i+1\sigma}+c^{\dagger}_{i+1\sigma}c_{i\sigma}\right)+
E_{f}\sum_{i\sigma}n^{f}_{i,\sigma} \nonumber 
\\ &+&V\sum_{i,\sigma}\left(c^{\dagger}_{i\sigma}f_{i\sigma}+f^{\dagger}_{i\sigma}c_{i\sigma}\right)+
U_{f}\sum_{i}n^{f}_{i\uparrow}n^{f}_{i\downarrow}\nonumber \\ 
&+& W\sum_{i,\sigma}\left(i-\frac{L+1}{2}\right)^{2}\left( n^{c}_{i,\sigma} +n^{f}_{i,\sigma} \right), 
\end{eqnarray}
\noindent where, $i$ varies along the sites of a one-dimensional lattice of size $L$, $c^{\dagger}_{i\sigma}(f^{\dagger}_{i\sigma})$ creates an atom 
at site $i$ with state $^{1}S_{0}$($^{3}P_{0}$) and with spin $\sigma= \uparrow, \downarrow$. The local number operator with spin $\sigma$ for the $g$ 
and $e$  atoms is $n^{c}_{i\sigma}=c^{\dagger}_{i\sigma}c_{i\sigma}$, and $n^{f}_{i\sigma}=f^{\dagger}_{i\sigma}f_{i\sigma}$, respectively. The 
strength of the hybridization is  $V=g_{l}\int d^{3}\textbf{r}\omega_{g}(\textbf{r})\omega_{e}(\textbf{r})$, where $\omega_{e,g}(\textbf{r})$ are 
real Wannier functions for each state and $g_{l}$ is an interaction parameter chosen with a suitable driving laser frequency and 
polarization~\cite{Zhong-FP17}. $E_{f}$ is the energy level of the $e$ atoms, and the local repulsion between them is 
$U_{f}=g_{ee}\int d^{3}\textbf{r}\omega^{4}_{e}(\textbf{r})$, where $g_{ee}$ is an interaction parameter. The hopping parameter is given by 
$t=-\int d^{3}\textbf{r}\omega_{g}(\textbf{r})(-\frac{\hbar}{2M}\bigtriangledown^{2}+V_{g}(\textbf{r}))\omega_{g}(\textbf{r}+\vec{\delta})$, where 
$V_{g}(\textbf{r})$ is an external potential, $M$ the atom's mass, and $\vec{\delta}$ is the nearest-neighbor vector. The quadratic potential 
term in the Hamiltonian was included in order to consider the confining potential due to the trap, and the strength parameter $W$ chosen must, in 
numerical simulations, be large enough to assure that there are no atoms at the edges of the chain. A schematic diagram of the model is shown 
in Fig. \ref{fig1}.\par
\begin{figure}[t!]
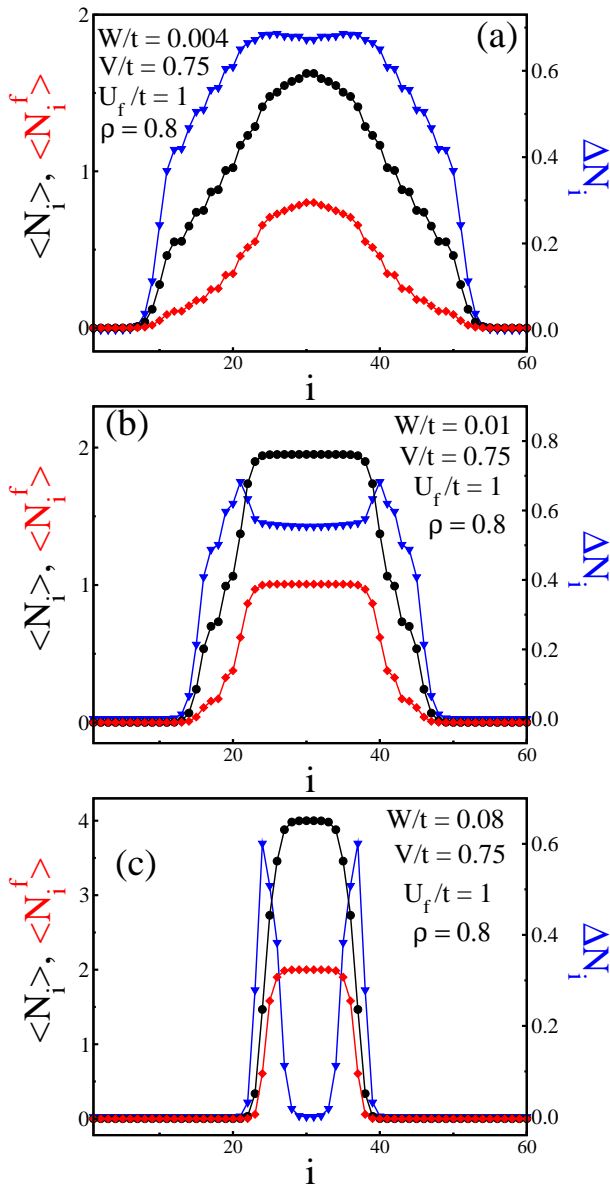

\begin{minipage}{19pc}
\includegraphics[width=19pc]{Fig2a.eps}
\end{minipage}
\hspace{5pc}%
\begin{minipage}{19pc}
\includegraphics[width=19pc]{Fig2b.eps}
\end{minipage}
\hspace{5pc}%
\begin{minipage}{19pc}
\includegraphics[width=19pc]{Fig2c.eps}
\caption{\label{fig2}Evolution of the local density profile and the variance as a function of the confining potential of strength $W$ for an optical 
lattice with $L=60$ sites, a global density of $\rho=0.8$, hybridization strength of $V/t=0.75$, and local repulsion $U_{f}/t=1$ and $E_f=-U_{f}/2$. From 
top to bottom, the confining potential increases as $W/t=0.004$ (a), $W/t=0.01$ (b), and  $W/t=0.08$ (c). The black data points correspond to 
the total number of atoms per site  ($\langle N_{i}\rangle$), the red data points correspond the number of localized atoms per site 
($\langle N^{f}_{i}\rangle$), and the blue data points show the variance along the lattice. The lines are visual guides.}
\end{minipage}\hspace{2pc}%
\end{figure}
Neglecting the confining potential in Hamiltonian (\ref{HPAM}) ($W_e=W_g=W=0$), we recover the one-dimensional periodic Anderson model (PAM), which 
is the standard model for studying the physics of heavy fermion materials. Despite the huge number of papers about the PAM, there are still many 
questions to be answered. In one dimension, we know that at half-filling the ground state is always an insulating spin liquid, where the magnetic 
correlations decay exponentially with the distance for all nonzero repulsion. At quarter filling, a metal-insulator transition from a 
paramagnetic metal to an insulator with a localized-band spin-density wave was found. For incommensurate densities between a quarter and half-filling, the 
ground state is always metallic, but the magnetic behavior is quite diverse; for instance, we found paramagnetic, ferromagnetic, incommensurate 
spin density wave (ISDW), and RKKY regions~\cite{Bertussi-PRB11}.\par
For alkaline-earth-metal atoms confined in optical lattices, other models related to the Hamiltomian (\ref{HPAM}) have been suggested; for instance, 
the model called ``$g$-$e$'' takes into account the kinetic energy of the $e$ atoms, a general local hybridization between the optical lattices, 
and a local repulsion between atoms in different optical lattices~\cite{Bois-PRB16}.

\section{\label{sec3}  Results} 

Our main goal in this paper is to explore the ground state of the periodic Anderson model with a harmonic potential given by the Hamiltonian 
(\ref{HPAM}), considering the same confining potential for the delocalized and localized atoms (see Fig. \ref{fig1}) and fixing a global density 
($\rho=N/L$, where $N$ is the total number of atoms, delocalized plus localized) in the system. Note that for a given global density, the values of 
the confining potential considered will be those for which all the particles are confined. We used the finite-size algorithm of the density matrix 
renormalization group method to find the ground state, keeping up to $600$ states per block and obtaining a discarded weight of around $10^{-5}$ or 
less. Our inhomogeneous problem converges very slowly, and we had to perform many sweeps (approx. 17) until convergence was reached. For this 
reason, we considered lattice sizes up to $L = 60$, but this size also corresponds to typical values in experimental setups. In the 
Hamiltonian (\ref{HPAM}), the hopping integral $t$ is set to 1, and we assume that the lattice constant is equal to 1.\par
An inhomogeneous distribution of the atoms is generated by the harmonic potentials, and we expected an accumulation of atoms in the middle of the 
trap, which will depend on the global density, the harmonic potential strength, the local hybridization, and local repulsion. Due to the 
inhomogeneous distribution of atoms in the system, we focused on local quantities in order to characterize the ground state of the system, such as the local 
number of localized atoms $\langle N^{f}_{i}\rangle=\langle \sum_{\sigma}n^{f}_{i,\sigma}\rangle$, the local total number of atoms 
$\langle N_{i}\rangle=\langle \sum_{\sigma}(n^{c}_{i,\sigma}+n^{f}_{i,\sigma})\rangle$, and the variance of the local total density per site 
$\Delta N_{i}=\langle N^{2}_{i}\rangle-\langle N_{i}\rangle^{2}$~\cite{Batrouni-PRL02,Rigol-PRA04}. The above quantities are shown in 
Fig. \ref{fig2} for a lattice with a global density of $\rho=0.8$, a hybridization strength of $V/t=0.75$, and local repulsion $U_{f}/t=1$ and 
$E_f=U_{f}/2$, for different values of the confining potential. In Fig. \ref{fig2} (a), a confining potential of $W/t=0.004$ was considered, and we note 
that at the ends of the lattice there are no atoms, i.e. the occupation is zero, showing that for this and larger strength potentials, all the atoms 
are confined. The occupation increases from zero after some sites and exhibits a monotonous increase as we approach the center of the trap as 
expected because of the harmonic potential. The local correlation between the optical lattices and the local repulsion between the localized atoms 
appear slight. Note that the occupation of the localized atoms ($\langle N^{f}_{i}\rangle$) is smaller than the occupation in the delocalized 
optical lattice. The ground state has a metallic region around the middle of the trap, and it is surrounded by band insulator regions with zero 
carriers. This picture is confirmed by the variance of the local density $\Delta N_{i}$, because there are no charge fluctuations at the ends and 
the variance is nonzero elsewhere. Increasing the confining potential up to $W/t=0.01$ (see Fig. \ref{fig2} (b)), we observed that the outer band 
insulator regions increase, showing that the system is more confined. Moving towards the middle of the trap, we see that both the total and localized 
number of atoms per site increase and reach constant values of $\langle N_{i}\rangle=2$ and $\langle N^{f}_{i}\rangle=1$, respectively. This means that 
around the middle of the trap, we have one atom at each site in both optical lattices, and they form a local singlet, establishing a spin liquid 
region for a global density $\rho=0.8$; therefore, the charge redistribution generated by the harmonic potential and the transitions between the 
optical lattices stimulated by the local repulsion and the hybridization compete to reach a region with local singlets for this low global density. 
Note that the spin liquid phase for the well-known homogeneous periodic Anderson model can only be reached when the global density is 
$\rho=2$~\cite{Bertussi-PRB11}. This means that harmonic potentials due to optical lattices open the door to obtaining spin liquid regions for different 
sets of parameters. The ground state 
in this case exhibits (from the end to the center) a band insulator region, a metallic region with $0<\langle N_{i}\rangle<2$, and a spin liquid domain. 
In the metallic region, we see that the total occupation $\langle N_{i}\rangle$ increases faster than $\langle N^{f}_{i}\rangle$ reflecting the effect 
of the local repulsion and the hybridization. The variance is zero in the band insulator regions and increases in the metallic regions, but in the spin 
liquid domain it remains constant, and this value is a local minimum, which indicates the presence of an incompressible, insulating region where the 
local compressibility $\kappa_i=\partial N_i/\partial \mu_i=\beta\Delta N_i$ drops to a small but finite value. A band insulator region with full 
sites around the middle of the trap is shown in Fig. \ref{fig2} (c) for a confining potential of $W/t=0.08$. In this case, two band insulators and a 
metallic region coexist in the ground state. This picture is corroborated by the variance, which is nonzero only in the metallic domains, showing 
that in the other regions there are no charge fluctuations. We can conclude that in the ground state of the periodic Anderson model with a 
harmonic potential, a spin liquid domain and metallic and band-insulator regions coexist, and the particular configuration of the ground state will 
depend on the value of the confining potential when the other parameters are fixed.\par
\begin{figure}[t!]
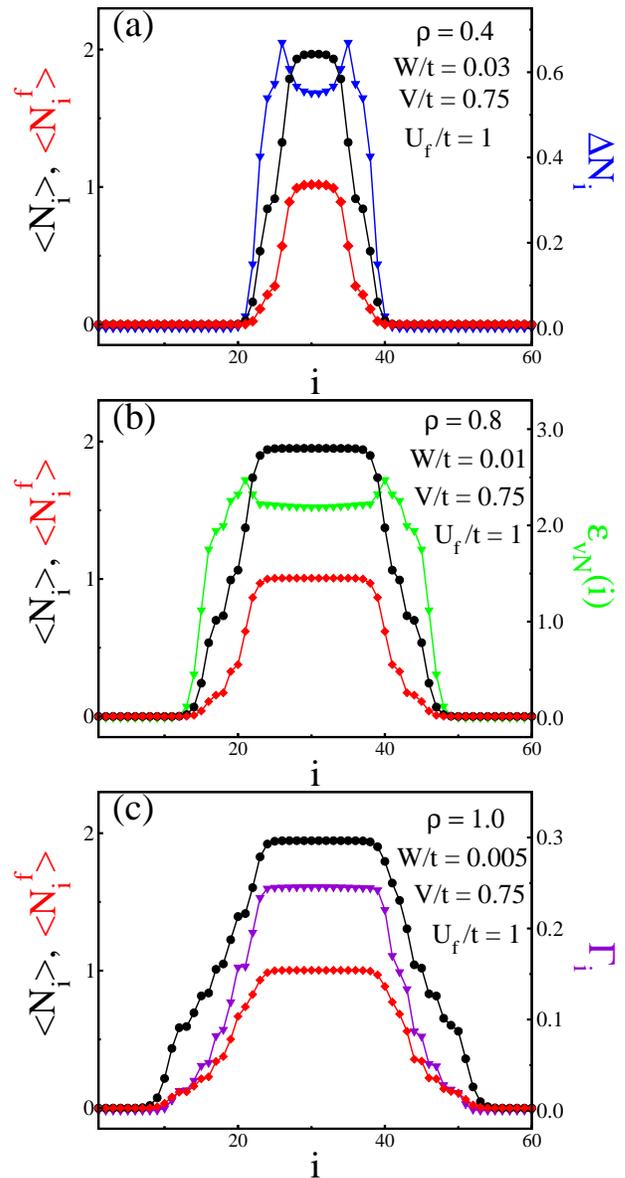

\begin{minipage}{19pc}
\includegraphics[width=19pc]{Fig3a.eps}
\end{minipage}
\hspace{5pc}%
\begin{minipage}{19pc}
\includegraphics[width=19pc]{Fig3b.eps}
\end{minipage}
\hspace{5pc}%
\begin{minipage}{19pc}
\includegraphics[width=19pc]{Fig3c.eps}
\caption{\label{fig3} Local density profile for three different global densities $\rho=0.4$, $0.8$, and $1.0$, the confining potential being 
$W/t=0.03$, $0.01$, and $0.005$, respectively. Here the hybridization strength is $V/t=0.75$, local repulsion $U_{f}/t=1$, and 
$E_f=-U_{f}/2$. The black data points correspond to the total number of atoms per site  ($\langle N_{i}\rangle$), the red data points 
correspond the number of localized atoms per site ($\langle N^{f}_{i}\rangle$). In the top panel we show the variance along the lattice (blue data 
points), while in the middle panel the local von Neumann entropy (green data points) is displayed, and finally in the bottom panel we show 
the local correlation between atoms in different optical lattices (violet data points). The lines are visual guides.}
\end{minipage}\hspace{2pc}%
\end{figure}
Kondo spin liquid domains surrounded by metallic regions and coexisting with band insulator regions were reported before exploring the ground state of 
the Kondo lattice model with a harmonic potential for the delocalized carriers~\cite{FossFeig-PRA10a,FossFeig-PRA10b,JSV-EPJB12a,JSV-PRA12}.\par 
In Fig. \ref{fig3}, we show the ground state density profile for three different global densities at a confining potential value, displaying a Kondo 
spin liquid domain around the middle of the trap. In particular, we chose $W/t=0.03$, $W/t=0.01$, and $W/t=0.005$ for the densities 
$\rho=0.4$, $\rho=0.8$, and $\rho=1.0$, respectively. We found that it is possible to obtain spin liquid domains for very low densities such as 
$\rho=0.4$ (see  Fig. \ref{fig3} (a)), for which the number of singlets are not large. Note that again the variance is constant around the middle of 
the trap, this being a local minimum, showing that this region is an insulator. In this case, in the ground state a band insulator, a metallic, and a spin 
liquid domain coexist. A larger spin liquid domain for a global density of $\rho=0.8$ is shown in Fig. \ref{fig3} (b), whose parameters coincide with 
those of Fig. \ref{fig2} (b), but now we show the local von Neumann entropy along the lattice, which is given by 
\begin{equation}\label{eq:SVN}
\varepsilon_{vN}(i)=-Tr(\rho_{i})log_{2}(\rho_{i}),
\end{equation}
\noindent where $\rho_i=Tr_{B} (|\Psi\rangle\langle\Psi|)$ is the density matrix of a single site located at $i$, the environment $B$ is composed of 
the remaining  $L-1$ sites, and $|\Psi\rangle$ is the ground state wavefunction. Nowadays, it is common to use the tools of quantum information theory 
to characterize quantum states~\cite{Amico-RMP08}, and in particular the local von Neumann entropy is very useful for studying inhomogeneous 
systems~\cite{Franca-PRL08,JSV-PB09}.\par 
In the band insulator regions with empty sites, the entanglement of an empty site with the rest of the lattice is zero; therefore, the local von Neumann 
entropy vanishes in the band insulator regions in the same way as the variance (see  Fig. \ref{fig3} (b)), and the entanglement increases in the metallic 
regions as expected, due to the quantum fluctuations. When local singlets are formed, the number of degrees of freedom decreases, and we expected that 
the entanglement would drop, but additionally the local von Neumann entropy remains constant and exhibits a plateau in the spin liquid domain. Comparing 
Fig. \ref{fig2} (b) and Fig. \ref{fig3} (b), we note that the local von Neumann entropy and the variance are constant in the insulator domains and 
vary in the metallic regions. It is clear that there is a one-to-one correspondence between the variance of the density and the von Neumann entropy; 
therefore, we can use the local von Neumann entropy to characterize the ground state of the periodic Anderson model with a harmonic potential. Also, 
this result leads to the possibility of estimating the entanglement in interacting systems through the measurement of the local variance using the 
high spatial resolution imaging microscopes developed recently for the observation of the local density distribution or the parity of the atom 
number on lattice sites~\cite{Gemelke-N09,Bakr-N09,Haller-NP15,Yamamoto-NJP16}.\par
\begin{figure}[t] 
\includegraphics[width=20pc]{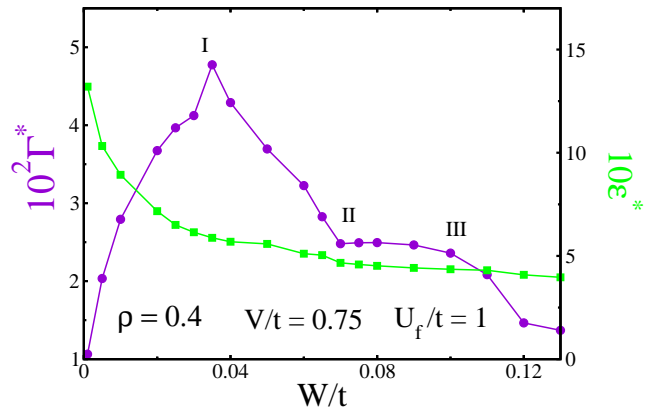} 
\caption{\label{fig4} The average spin-spin correlation (violet points) and von Neumann entropy (green points) as a function of the confining 
potential $W/t$. The parameters involved are $\rho=0.4$, $V/t=0.75$, $U_{f}/t=1$, and $E_f=U_{f}/2$. I, II, and III mark where the transitions 
take place. The lines are visual guides.} 
\end{figure} 
As the global density increases, the confining potential needed to obtain a spin liquid domain around the middle of the trap decreases, as can be 
seen in Fig. \ref{fig3}. In particular, for $\rho=1.0$ (see Fig. \ref{fig3} (c)), we show that another local quantity is useful for characterizing 
the ground state of Hamiltonian (\ref{HPAM}), which is the on-site spin-spin correlation between atoms in different optical lattices, given by 
\begin{equation}\label{eq:Scf}
\Gamma_i= |\langle\Psi |\mathbf{S}_{i}^{c}\cdot \mathbf{S}_{i}^{f}|\Psi\rangle|,
\end{equation}
\noindent where 
$\mathbf{S}_{i}^{\eta}=\frac{1}{2}\sum_{\alpha\beta}\hat{\eta}^{\dagger }_{i,\alpha }\boldsymbol{\sigma}_{\alpha\beta}\hat{\eta}_{i,\beta }$ 
is the spin operator of the localized ($\eta=f$) or delocalized ($\eta=c$) atoms, $\boldsymbol{\sigma}$ being the vector of Pauli matrices.\par
At empty sites, there is no correlation; therefore, the spin-spin correlation vanishes in the band insulator regions, and then the correlation increases 
monotonously in the metallic regions, reaching the maximum value around the middle of the trap, where a singlet is formed at each site. The evolution 
of the spin-spin correlation through the lattice and the above discussion of figures \ref{fig3} (a) and \ref{fig3} (b) allow us to conclude that 
between local variance, the local von Neumann entropy, and the on-site spin-spin correlation there is a one-to-one correspondence.\par 
So far we have shown that band insulator, metallic, and Kondo spin liquid regions coexist in the ground state of the periodic Anderson model with a 
harmonic potential, but the particular configuration of the ground state will depend on the set of parameters chosen. Thus a change of state can be 
observed when we vary some parameters while the others are fixed. Since our system is inhomogeneous, we use local quantities to characterize the ground 
state, which suggests that the evolution of the average of the local quantities can be useful for estimating the critical values for which the system 
changes its state. The average per-site entropy $\varepsilon^*=\tfrac{1}{L}\sum_i \varepsilon_{vN}(i)$ and its derivative were successfully used to 
determine critical state points in inhomogeneous Hubbard and Kondo lattice models~\cite{Franca-PRL08,JSV-PRA12}. Fixing the global density, the 
hybridization, the local energy of localized atoms, and the local repulsion, we show the evolution of the average per-site entropy with the confining 
potential in Fig. \ref{fig4}. The atoms will be confined if the harmonic potential strength is greater than $W/t=0.0008$. Starting at this value, the ground 
state is composed of outer band insulator regions and a central metallic domain, and the quantum entanglement will be finite, but as the confining 
potential increases, the system will be more confined, the outer band insulator regions grow, and the entanglement diminishes, as we can see in the 
figure. As the confining potential increases, we expected that a Kondo spin liquid domain would appear around the center of the lattice, which should 
mean a drop in the value of $\varepsilon^*$. However, we do not see that, and the evolution of the average per-site entropy is almost monotonous. In 
the derivative of the average per-site entropy, we obtain several extremes (maximum or minimum), but only one has physical meaning, indicating the 
true state transition from a ground state with a spin liquid to a metallic domain around the middle of trap. We believe that this a consequence of 
the small number of states retained in the DMRG process ($m=600$). With larger number of states, $\varepsilon^*$ and their derivative would give better 
results, as in the Kondo lattice model with a harmonic potential~\cite{JSV-PRA12}.\par
\begin{figure}[t!]
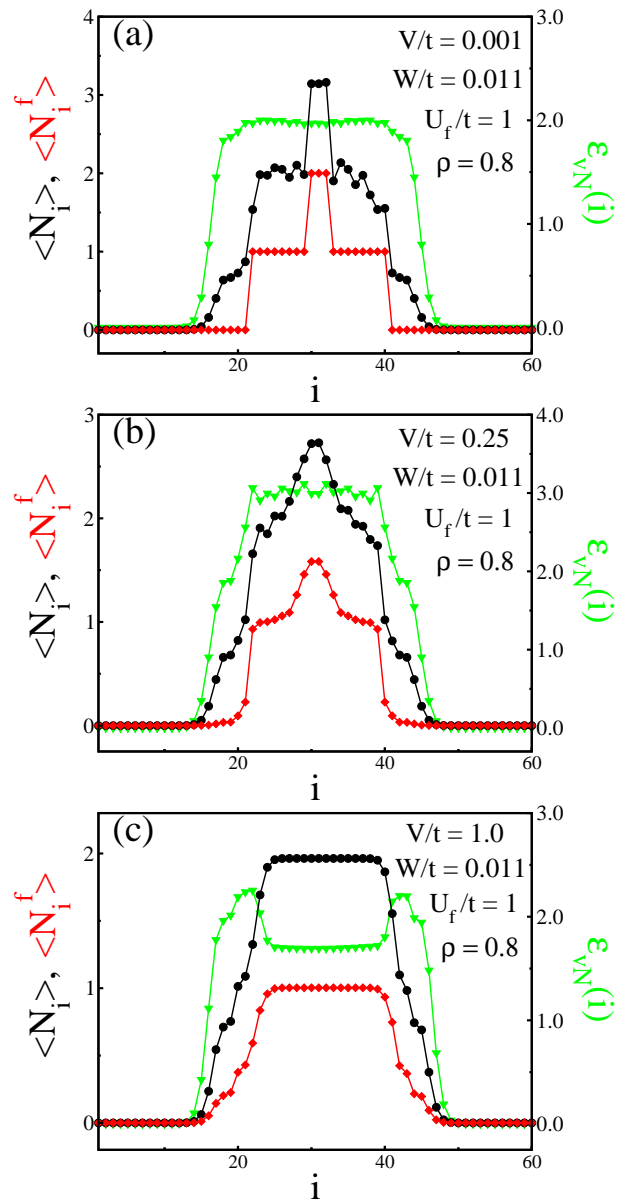

\begin{minipage}{19pc}
\includegraphics[width=19pc]{Fig5a.eps}
\end{minipage}
\hspace{5pc}%
\begin{minipage}{19pc}
\includegraphics[width=19pc]{Fig5b.eps}
\end{minipage}
\hspace{5pc}%
\begin{minipage}{19pc}
\includegraphics[width=19pc]{Fig5c.eps}
\caption{\label{fig5} Evolution of the local density profile as a function of the hybridization $V$ for an optical lattice with 
$L=60$ sites, a global density of $\rho=0.8$, a confining potential of $W/t=0.01$, local repulsion $U_{f}/t=1$, and $E_f=U_{f}/2$. From 
top to bottom, the hybridization increases as $V/t=0.001$ (a), $V/t=0.25$ (b), and  $V/t=1.0$ (c). The black data points correspond to 
the total number of atoms per site  ($\langle N_{i}\rangle$), the red data points correspond the number of localized atoms per site 
($\langle N^{f}_{i}\rangle$), and the green data points show the local von Neumann entropy along the lattice. The lines are visual guides.}
\end{minipage}\hspace{2pc}%
\end{figure}

With the goal of estimating the critical points, we decided to try using the average of the on-site spin-spin correlation 
$\Gamma^*=\tfrac{1}{L}\sum_i \Gamma_i$, which is also shown in Fig. \ref{fig4}. For small values of the confining potential, the ground state has 
small outer band insulator regions and a big metallic region. The average occupation in the localized optical lattice is lower than one; therefore, 
the average spin-spin correlation is small, but it increases as the harmonic potential grows, because the outer band insulator region increases and the 
average number of atoms around the center grows. Hence the average spin-spin correlation is larger. A quantum change of state happens at 
$W/t=0.035$. The average spin-spin correlation reaches a maximum, and the ground state has a Kondo spin liquid around the middle of the trap. A 
subsequent increase in the confining potential decreases the Kondo spin liquid domain and the metallic regions and increases the outer band insulator 
regions. That is why the average spin-spin correlation decreases despite having a spin liquid domain. The ground state changes at $W/t=0.07$, and the 
central spin liquid domain disappears, being replaced by a metallic region. Now the metallic domain decreases, the outer band insulator increases, and the 
double occupation compete with each other, leading to a constant value of the average spin-spin correlation. Finally, a quantum state transition takes place at 
$W/t=0.10$ from a state with a metallic domain in the middle of the trap to a state with a band insulator of full sites around the center of the 
lattice. Due to the local double occupancy in both optical lattices in the central band insulator domain, the average spin-spin correlation decreases. 
In spite of the fact that the average  per-site entropy does not give us information about the change of state, the average spin-spin correlation showed us three 
quantum state transitions.\par
\begin{figure}[t!]
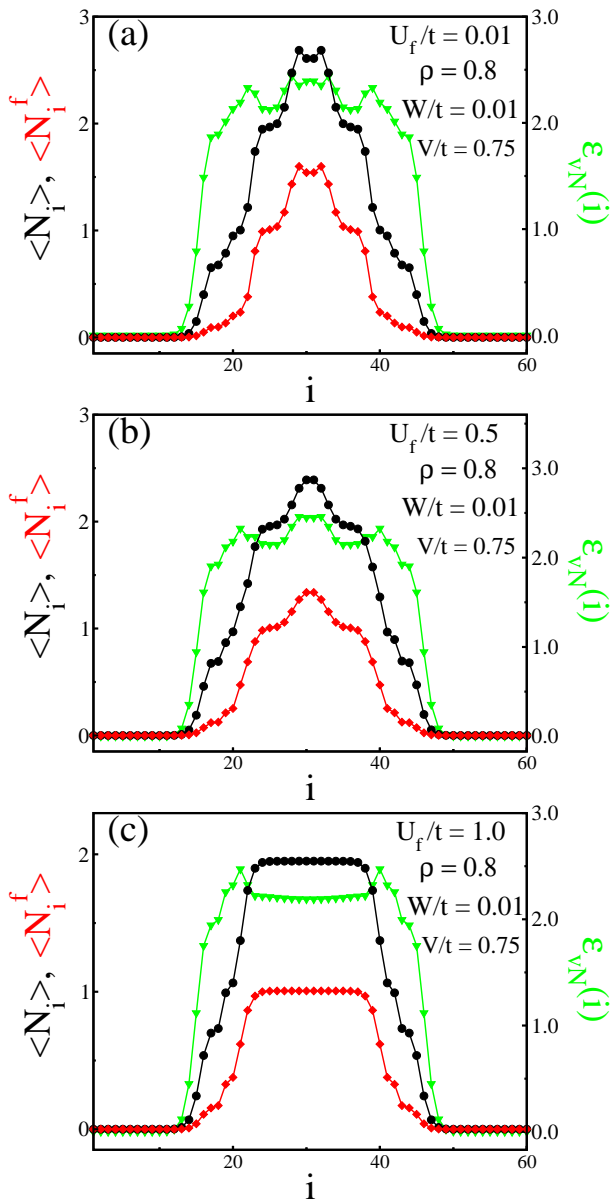

\begin{minipage}{19pc}
\includegraphics[width=19pc]{Fig6a.eps}
\end{minipage}
\hspace{5pc}%
\begin{minipage}{19pc}
\includegraphics[width=19pc]{Fig6b.eps}
\end{minipage}
\hspace{5pc}%
\begin{minipage}{19pc}
\includegraphics[width=19pc]{Fig6c.eps}
\caption{\label{fig6} Local density profile for three different values of the local repulsion, $U_{f}/t=0.001$, $0.5$, and $1.0$. The other parameters 
are: $\rho=0.8$, $W/t=0.01$, $V/t=0.75$, $E_f=U_{f}/2$, and $L=60$. The black data points correspond to 
the total number of atoms per site  ($\langle N_{i}\rangle$), the red data points correspond the number of localized atoms per site 
($\langle N^{f}_{i}\rangle$), and the green data points show the local von Neumann entropy along the lattice. The lines are visual guides.}
\end{minipage}\hspace{2pc}%
\end{figure}
The evolution of the density profile for three different values of the hybridization for the periodic Anderson model with a harmonic potential is 
shown in Fig. \ref{fig5}. In this figure, we consider a global density of $\rho=0.8$, a confining potential of $W/t=0.01$, and a local repulsion of $U_{f}/t=1$ 
and $E_f=U_{f}/2$. For small values of the hybridization, we expected that more atoms would go to the delocalized lattice in order to minimize the system energy.  
Due to this, we observed that the $e$ atoms are more localized than the $g$ ones, generating a region without any localized atom and a nonzero 
occupation of the delocalized ones, which characterize a phase separation domain, this being a new fact not found before in systems of carriers 
confined in a harmonic potential (see Fig. \ref{fig5} (a)). The phase separation domain persists for finite values of the hybridization 
(Fig. \ref{fig5} (b)) and disappears for larger values as the quantum fluctuations between the lattices increases (Fig. \ref{fig5} (c)). Returning 
to the case of small hybridization values, we note that due to the local repulsion and in order to minimize the energy, the localized atoms form a small doubly 
occupied region around the center of the trap and long plateaus surrounding the latter, with one atom per site, whereas the occupation of the 
delocalized lattice in these regions fluctuates around one (see Fig. \ref{fig5} (a)). Increasing the hybridization causes the quantum 
fluctuations to grow, the doubly occupied sites to disappear, and the distribution of the localized atoms along the lattice  to be smoothed, conserving the 
plateaus. Finally, in Fig. \ref{fig5} (c), we see that for $V/t=1.0$ the phase separation domain disappears, and only one plateau is displayed, 
corresponding to the formation of the Kondo spin liquid state. Therefore, we show that it is possible to tune the latter state by changing the 
hybridization.\par 
Throughout this report, we have explored the ground state of Hamiltonian (\ref{HPAM}) as a function of the confining potential, the global density, and 
the hybridization, but we have not yet explored the behavior as a function of the local repulsion, which is shown in Fig. \ref{fig6} for $U_{f}/t=0.001$, 
$0.5$, and $1.0$. The parameters that remain fixed in this figure are: $\rho=0.8$, $W/t=0.01$, $V/t=0.75$, and $E_f=U_{f}/2$. In Fig. \ref{fig6} (a), 
we observe that outer band insulator regions and a central metallic one coexist in the ground state, because the atoms try to distribute themselves between  
both lattices, except near the band insulator regions. Although a decrease is observed in the variance for a few sites, this does not remain constant, and also 
a single singlet is observed; therefore, this small number of sites does not constitute a Kondo spin liquid domain, and we talk about a central 
metallic region. However, this allows us to note that the hybridization is more important than the local repulsion ($U_{f}/t=0.001$) for the generation 
of the Kondo spin liquid domains. As the local repulsion increases, more atoms prefer to go to the delocalized lattice and the distribution of atoms 
in both lattices is smoothed, showing two small Kondo spin liquid domains surrounding a central metallic region for $U_{f}/t=0.5$ (Fig. \ref{fig6} (b)). 
This process will continue until the central metallic region disappears and outer band insulator regions and central Kondo spin liquid domain coexist 
in the ground state, as shown in Fig. \ref{fig6} (c) for $U_{f}/t=1.0$. The width of the Kondo spin liquid domain diminishes as the local repulsion 
increases until the total number of atoms per site is lower than two and the ground state is composed of the outer band insulator regions and a 
 central metallic domain.\par 

\section{\label{sec4} Conclusions}
We investigated under what conditions it is possible to obtain Kondo spin liquid features with alkaline-earth-metal atoms confined in 
one-dimensional optical lattices with an effective hybridization. We modeled the above system using a one-dimensional periodic Anderson model plus a
harmonic potential for both type of carriers and the well-known density matrix renormalization group algorithm was used to explore the ground state as 
a function of the parameters. The inhomogeneity of the system generates a redistribution of atoms in both optical lattices, which is reinforced by the 
quantum fluctuations induced by the local hybridization, leading to a ground state where diverse metallic and insulator domains coexist. We found that 
in this inhomogeneous periodic Anderson model, it is possible to obtain Kondo spin liquid domains, i.e. the formation of local singlets between atoms 
in different optical lattices on a portion of the chain. These Kondo spin liquid domains can be obtained for low global densities and tuned by varying 
the confining potential, the hybridization, and the local repulsion. We used local quantities such as a variance of the total number of atoms, the 
spin-spin correlation between atoms in different optical lattices and the von Neumann entropy to characterize the ground state, finding a
one-to-one correspondence between these quantities, which opens the possibility of indirectly estimating the entanglement of an interacting system in 
cold atom setups through the measurement of variance, using high spatial resolution imaging microscopes.\par 
We obtained that band insulator regions with empty or full sites, metallic regions, a phase separation region, and Kondo spin liquid domains coexist 
in the ground state, and the particular configuration will depend on the set of parameters considered. The critical points corresponding to the change of 
state can be determined by the average of the on-site spin-spin correlation function and perhaps by the average per-site entropy, if a larger number 
of states is retained in the DMRG algorithm.\par 

\section*{Acknowledgments}
The authors are thankful for the support of Departamento Administrativo de Ciencia, Tecnolog\'{\i}a e Innovaci\'on (COLCIENCIAS) 
(Grant No. FP44842-057-2015). J.S.-V. is grateful for the hospitality of the ICTP, where part of this work was done.

\bibliography{Bibliografia}

\end{document}